\documentclass[journal=jacsat,manuscript=article]{achemso}

\usepackage{amsmath,amsfonts,amsthm,amssymb,array,bm,dcolumn,enumerate,flafter,float,graphics,latexsym,makeidx,multirow,natbib,verbatim,xfrac}
\usepackage{graphicx}
\usepackage[english]{babel}
\usepackage[mathscr]{euscript}
\usepackage[utf8]{inputenc}
\usepackage[usenames]{color}
\usepackage[dvipsnames]{xcolor}
\usepackage{chemfig}
\usepackage[pdftex,colorlinks=true,linkcolor=blue,urlcolor=blue,citecolor=blue]{hyperref}

\usepackage[version=3]{mhchem} % Formula subscripts using \ce{}

\author{Victor Lopez-Richard}
\email{vlopez@df.ufscar.br}
\affiliation{Departamento de Física, Universidade Federal de São Carlos, 13565-905, São Carlos, SP, Brazil}

\author{Igor Ricardo Filgueira e Silva}
\affiliation{Departamento de Física, Universidade Federal de São Carlos, 13565-905, São Carlos, SP, Brazil}

\author{Alessandra Ames}
\affiliation{Departamento de Física, Universidade Federal de São Carlos, 13565-905, São Carlos, SP, Brazil}

\author{Frederico B. Sousa}
\affiliation{Departamento de Física, Universidade Federal de São Carlos, 13565-905, São Carlos, SP, Brazil}

\author{Marcio Daldin Teodoro}
\affiliation{Departamento de Física, Universidade Federal de São Carlos, 13565-905, São Carlos, SP, Brazil}

\author{Ingrid David Barcelos}
\affiliation{Brazilian Synchrotron Light Laboratory (LNLS), Brazilian Center for Research in Energy and Materials (CNPEM), Campinas, SP 13083-100, Brazil}

\author{Raphaela de Oliveira}
\affiliation{Brazilian Synchrotron Light Laboratory (LNLS), Brazilian Center for Research in Energy and Materials (CNPEM), Campinas, SP 13083-100, Brazil}

\author{Alisson Ronieri Cadore}
\affiliation{Brazilian Nanotechnology National Laboratory (LNNano), Brazilian Center for Research in Energy and Materials (CNPEM), Campinas, SP 13083-100, Brazil}

\title[An \textsf{achemso} demo]
  {The Emergence of Mem-Emitters} 
  
\abbreviations{IR,NMR,UV}

\begin{document}

%\begin{tocentry}

%\includegraphics{Diagrama.pdf}
%\caption{Artistic representation of the light emission memory loop for excitons and trions  (Image partially generated using OpenArt.ai)}
%\end{tocentry}

\begin{abstract}

The advent of memristors and resistive switching has transformed solid state physics, enabling advanced applications such as neuromorphic computing. Inspired by these developments, we introduce the concept of Mem-emitters, devices that manipulate light emission properties of semiconductors to achieve memory functionalities. Mem-emitters, influenced by past exposure to stimuli, offer a new approach to optoelectronic computing with potential for enhanced speed, efficiency, and integration. This study explores the unique properties of transition metal dichalcogenides-based heterostructures as a promising platform for Mem-emitter functionalities due to their atomic-scale thickness, tunable electronic properties, and strong light-matter interaction. By distinguishing between population-driven and transition rate-driven Mem-emitters, we highlight their potential for various applications, including optoelectronic switches, variable light sources, and advanced communication systems. Understanding these mechanisms paves the way for innovative technologies in memory and computation, offering insights into the intrinsic dynamics of complex systems.

\end{abstract}


In recent decades, the field of solid state physics has been revolutionized by the emergence of memristors and the phenomenon of resistive switching~\cite{DiVentra2009}. First theorized in the 1970s,~\cite{Chua1971} memristors are circuit elements whose resistance can be altered based on their history. The underlying mechanism for resistive switching involves changes in a material's conductance in response to electrical or other stimuli that interfere with internal state variables.~\cite{LopezRichard2022} This capability to modulate resistance has unlocked exciting possibilities for neuromorphic computing,~\cite{Kumar2022} and other advanced applications in the realm of miniaturized solid state devices characterized by greater efficiency, density, and a variety of functionalities.~\cite{Rios2016,Gadelha2019,Lanza2022,Milano2022}

Inspired by the concept of resistive switching in memristors, we propose a paradigm shift in optoelectronics. This approach focuses on manipulating the light emission properties of semiconductors to achieve memory functionalities using through methods such as electromagnetic field modulation and proximity effects. We will refer to these devices as Mem-emitters where past exposure to light or other stimuli controls the characteristics of their emitted light. This enables a memory scheme where distinct light emission patterns represent different information states.

The exploration of Mem-emitters holds potential for the development of light-based memory devices. This technology could open up possibilities in optoelectronic computing, paving the way for advancements in speed, efficiency, and seamless integration with existing technologies.

The main observables in the optical response of the Mem-emitter are the emission intensity and its energy position. To better frame, qualify, and quantify the light emission intensity, we can reduce it to its fundamental components: the population of the initial state, $N$, and the optical transition rate, $1/\tau$. The intensity $Q$ can be expressed as 
\begin{equation}
    Q=\frac{N}{\tau}.
    \label{q}
\end{equation}
In the simplest approximation, the optical transition rate can be expressed as 
\begin{equation}
    \frac{1}{\tau}=K_{f,i}\cdot D\left( \Delta E_{f,i} - \hbar \omega\right),
    \label{tau}
\end{equation}
where $K_{f,i}$ is proportional to the optical transition matrix elements between the initial (i) and final state (f); and $D\left( \Delta E_{f,i} - \hbar \omega\right)$ represents the energy conservation factor, which peaks when the photon energy ($\hbar \omega$) approaches the  energy difference between these two states ($\Delta E_{f,i}$), i.e, $\hbar \omega \simeq \Delta E_{f,i}$.

\begin{figure}[!htb]
    \centering
   \includegraphics[width=0.8\textwidth]{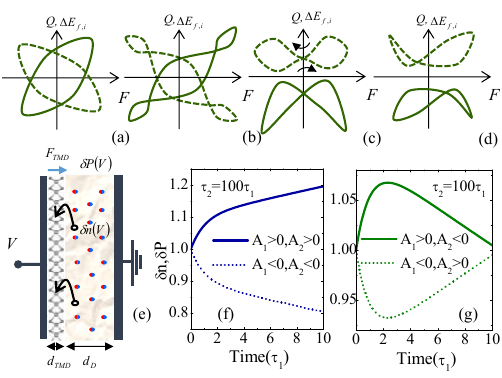}
    \caption{(a)-(d) Illustrative depiction of potential hysteresis loops of a Mem-emitter under periodic stimulus, showcasing various possible topological shapes: solid and dashed lines highlight contrasting loop configurations, and the arrows in panel (c) indicate a pinched, non-crossing loop. (e) Representation of a Mem-emitter consisting of a heterostructure based on a TMD monolayer of width $d_{TMD}$ on its dielectric substrate of width $d_{D}$, connected to two electrodes. The nonequilibrium carriers activated in the substrate and reaching the TMD layer are represented by $\delta n$ along with the electric polarization fluctuation, $\delta P$, also occurring within the substrate. Time resolved dynamics of either carrier or polarization fluctuations in the presence of two independent mechanisms with (f) monotonic and (g) non-monotonic transients}
    \label{fig1}
\end{figure}

For the mem-emitting ability to emerge, the functions in Eqs.~\ref{q} and~\ref{tau} must depend not only on the external electromagnetic fields, $\Vec{F}$, but also on a set of internal state variables, denoted as a vector of factors, $\Vec{x}$. These internal state variables may include the density of activated or trapped carriers, internal electric polarization, redox reaction rates at the interfaces, filament formation rates, and other dynamic factors. Note that, in the presence of magnetic interactions, the spin degree of freedom must be considered. If the external fields vary, the time evolution of each emission intensity component: $K_{f,i}(\Vec{F},\Vec{x},t)$, $\Delta E_{f,i}(\Vec{F},\Vec{x},t)$, and $N(\Vec{F},\Vec{x},t)$ must be affected by dynamic relationships, that can be reduced in general terms to
\begin{equation}
    \frac{d \Vec{x} }{dt}=f \left( \Vec{F},\Vec{x},t \right).
    \label{alf}
\end{equation}
This dependence ensures that the light emission intensity, $Q$, and energy position are influenced by both external fields and the internal state of the system, allowing for an adaptive emission behavior. Given that the optical emission rate of the Mem-emitter at any given time is influenced by the entire history of the external field evolution, its behavior is inherently non-Markovian. The future state depends not only on the present state but also on the sequence of past states. This characteristic allows Mem-emitters to be used in applications that require memory of past conditions, such as in neuromorphic computing and adaptive circuits. A key signature of this behavior is the appearance of stable hysteresis in either (or both) the emission intensity and energy position under periodic external driving, as depicted in Fig.~\ref{fig1}.

The conditions under which these hysteresis loops appear and evolve by tuning external factors, along with their symmetry and the number and positions of intersection points, are crucial elements for mapping the internal processes that generate them. Although a comprehensive classification of the topology of these shapes has yet to be presented, we can identify several classes based on specific criteria. According to the presence of an opening around the origin at $F=0$, loops can be classified as (i) open loops (a,b,d), (ii) x-cross loops [3 (c)-solid line], or (iii) pinched non-crossing loops [3 (c)-dashed line]. Based on symmetry, loops can be categorized as (i) even-symmetric loops (c), (ii) odd-symmetric loops (a,b), or (iii) asymmetric loops (d). Loops can also be classified according to the number of crossings: zero [1(a)], one [1(c) and 1(d)], two [1(b)], and more. Furthermore, the direction of the loop provides another classification criterion: (i) clockwise or (ii) counter-clockwise loops. This detailed classification framework helps in understanding the diverse types of hysteresis behaviors and the underlying physical mechanisms, providing insights into the internal dynamics of the systems exhibiting these loops. Specific cases will be discussed subsequently.

Another aspect to be explored in the behavior of Mem-emitters is the contrast between population-driven and transition-rate-driven Mem-emitters. Population-driven Mem-emitters are characterized by their modulation through fluctuations in population density, $N$. In these devices, the primary mechanism influencing light emission is the variation in the number of charge carriers or emitters. This can be achieved through charge injection or extraction by adjusting the external electric field, which directly affects the population of excited states capable of emitting photons. This type of Mem-emitter is particularly suitable for applications where the control of the number of active emitters and the emission intensity, $Q$, is crucial, such as in optoelectronic switches or variable light sources. The response time of these devices is typically governed by the rates of charge carrier injection and recombination~\cite{Pfenning2015}.

On the other hand, transition rate-driven Mem-emitters are primarily modulated through the tuning of the optical transition rate ($1/\tau$) with electromagnetic fields. In these devices, the focus is on altering the probability of radiative transitions between energy levels rather than the population density. The intensity of the emission is controlled by tuning the wavefunctions involved in the transition rate between states, which can be influenced by external electromagnetic fields and internal variables. This approach is ideal for applications requiring precise control over emission timing and intensity, such as in advanced communication systems and high-precision sensors. The response time in transition rate-driven Mem-emitters is influenced by the dynamics of the transition rates, which can be fine-tuned for specific applications. It is important to note that population and transition rate-driven mechanisms can coexist and might interfere in many systems and devices during the measurements.

Mem-emitters built on functional TMD heterostructures
and two-dimensional (2D) semiconductors, in general, present a highly promising platform for exploring Mem-emitter functionalities due to several unique and advantageous properties~\cite{Manzeli2017}. Their atomic-scale thickness allows for extreme miniaturization of devices and results in a high surface-to-volume ratio, which enhances the interaction of the material with external fields and surrounding environments.~\cite{Rezende2019,Xue2022} The reduced dimensionality also facilitates better control over the material's electronic and optical properties, making them highly sensitive to external stimuli such as proximity effects (of electrochemical nature)~\cite{Meneses2018,Cabral2018,Buscema2014} and high local electric fields.~\cite{Shi2019,Gadelha2020} We should note that memory traces have already been observed in the optical response of of TMDs-based devices,~\cite{Newaz2013,Roch2018,Choi2024} but they have not yet been explored or characterized as the significant feature they evidently are.

Many TMDs, such as MoS$_2$, WS$_2$, and WSe$_2$, exhibit a direct bandgap in their monolayer form, which is crucial for efficient light emission~\cite{Chaves2020,Cadore2024}. The bandgap of these materials can be tuned by applying strain, electric fields, or by stacking different layers, enabling precise control over their optical properties~\cite{Chaves2020,Sousa2024}. This tunability is essential for developing transition rate-driven Mem-emitters where controlling the emission wavelength and intensity is key. They also exhibit exceptionally strong light-matter interaction due to their reduced dimensionality and high exciton binding energy~\cite{Wang2018,Chaves2020}. This results in enhanced optical absorption and emission efficiencies, making them ideal for optoelectronic applications~\cite{Tan2020}. Additionally, 2D TMDs possess well-resolved excitonic complex emission lines, such as neutral excitons and trions.~\cite{Plechinger2016,Prando2021,Timmer2024} These distinct emission lines offer a precise platform for probing mem-emitting abilities, as they provide clear, identifiable markers for studying the dynamics of light emission and its modulation under various external stimuli. The ability to resolve and manipulate these excitonic features enhances the potential for developing sophisticated and highly controllable Mem-emitter devices. Additionally, 2D semiconductors can be easily integrated with other materials, including traditional semiconductors, insulators, and metals, to form heterostructures and hybrid devices.~\cite{Lemme2022,Barcelos2023} This compatibility facilitates the design of complex devices with tailored properties, combining the advantages of different material systems to optimize performance for specific applications.

To exemplify our Mem-emitter concept, we will use a device consisting of a 2D TMD monolayer placed on a dielectric substrate and connected to two electrodes, as depicted in Figure~\ref{fig1} (e). Note that this heterostructure also serves as the gate setup for a single-layer TMD transistor~\cite{Radisavljevic2011,Gadelha2019,Nasiruddin2024}. In this capacitor configuration, under an applied bias, the electric field within the TMD monolayer, $F_{TMD}$, is related to the applied voltage, $V$, and the built-in polarization in the substrate, $P$, by the following expression
\begin{equation}
    F_{TMD}=\frac{V+ \frac{P(V)}{\epsilon_0}d_D   }{d_D\left( 1+\chi_{TMD}  \right) + d_{TMD}},
    \label{ftmd1}
\end{equation}
where $\epsilon_0$ is the vacuum permittivity, $d_{TMD}$  and $d_D$  are the thicknesses of the TMD monolayer and the dielectric layer, respectively; $\chi_{TMD}$ is the dielectric susceptibility of the monolayer. Let’s assume that the polarization of the substrate can be segmented into two contributions: one that varies instantaneously with the local field in the substrate region, $F_D$, as $P^{(0)}=\chi_{D} \epsilon_0 F_D$, being $\chi_{D}$ the dielectric susceptibility of the substrate, and a fluctuation, $\delta P$, due to the leakeage or generation of non-equilibrium dipoles. Thus, the total polarization can be expressed as
\begin{equation}
   P=P^{(0)}+\delta P.
    \label{p}
\end{equation}

In that case, the local field at the TMD monolayer transforms to
\begin{equation}
    F_{TMD}=\left[ V+ \frac{\delta P(V)}{\epsilon_0}d_D \right] \frac{\left( 1+\chi_{D}  \right)}{d_D\left( 1+\chi_{TMD}  \right) + d_{TMD} \left( 1+\chi_{D}  \right)},
    \label{ftmd2}
\end{equation}
which, under the condition $d_{TMD}<< d_D$, simplifies to 
\begin{equation}
    F_{TMD}=\left[ \frac{V}{d_D}+ \frac{\delta P(V)}{\epsilon_0} \right] \frac{\left( 1+\chi_{D}  \right)}{\left( 1+\chi_{TMD}  \right)},
    \label{ftmd3}
\end{equation}
Various concomitant processes of carrier trapping can contribute to the polarization fluctuation, $\delta P=\sum_j \delta P_j$.  We will consider these processes to be independent, each with its own relaxation time and each component $\delta P_j$ will define an internal state variable, as described in Eq.~\ref{alf} for our Mem-emitter definition. Under this approximation, the fluctuation for each process j can be described by
\begin{equation}
   \frac{d \delta P_j}{dt}=- \frac{\delta P_j}{\tau_j}+g_j^{(p)}(V),
    \label{dp}
\end{equation}
where $\tau_j$ is the relaxation time and $g_j^{(p)}(V)$ is the polarization transfer function dependent on the applied voltage $V$ described in detail in Ref.~\citenum{SilvaIgor2024}. The sign of the transfer function determines whether the process is of a generative or depletive nature. Additionally, the trapping and release of carriers lead to fluctuations in the available extra charges, $\delta n=\sum_i \delta n_i$, resulting from the independent generation of nonequilibrium channels. Each of these channels leads to charge fluctuations, $\delta n_i$, adding new components to our internal state variable vector (Eq.~\ref{alf}), which, in anaology to Eq.~\ref{dp}, can be described using the relaxation time approximation
\begin{equation}
   \frac{d \delta n_i}{dt}=- \frac{\delta n_i}{\tau_i}+g_i^{(n)}(V),
    \label{dn1}
\end{equation}
where $g_j^{(n)}(V)$ is the non-equilibrium carrier transfer function for each independent channel i that can be found defined in Ref.~\citenum{LopezRichard2022} in terms of microscopic elements. The most strighforward way to prove the coexistence of non-equilibrium processes with contrasting relaxation times is the analysis of the time resolved transientes of the observables under constant bias, $V_0$. In such conditions, using the charge fluctuation as reference, the solution is
\begin{equation}
   \delta n(V_0)= \sum_i \left[ g_i^{(n)}(V_0)\tau_i +A_i \exp\left(-\frac{t}{\tau_i}  \right) \right],
    \label{dn2}
\end{equation}
with $A_i=\delta n_i(0)-g_i^{(n)} (V_0 ) \tau_i$, where $\delta n_i(0)$ is the initial condition when the pulse $V_0$ is applied. The solution for the polarization fluctuation would be analogous. This equation highlights how different relaxation times manifest in the transient response of the system. Some results are illustrated in Figures~\ref{fig1} (f) and~\ref{fig1} (g). Figure~\ref{fig1} (f) shows monotonic trends, while Figure~\ref{fig1} (g) displays non-monotonic trends for two concurrent processes with contrasting relaxation times differing by two orders of magnitude. The monotonicity is determined by the parameters $A_i$, whose sign depends on the initial condition. Thus, no direct conclusion on the character (sign) of the transfer functions, $g_i^{(n)} (V_0 )$ or $g_j^{(p)} (V_0 )$, can be extracted from these trends in case the observables are related to either $\delta n$ or $\delta P$, respectively.

\begin{figure}[!htb]
    \centering
   \includegraphics[width=1\textwidth]{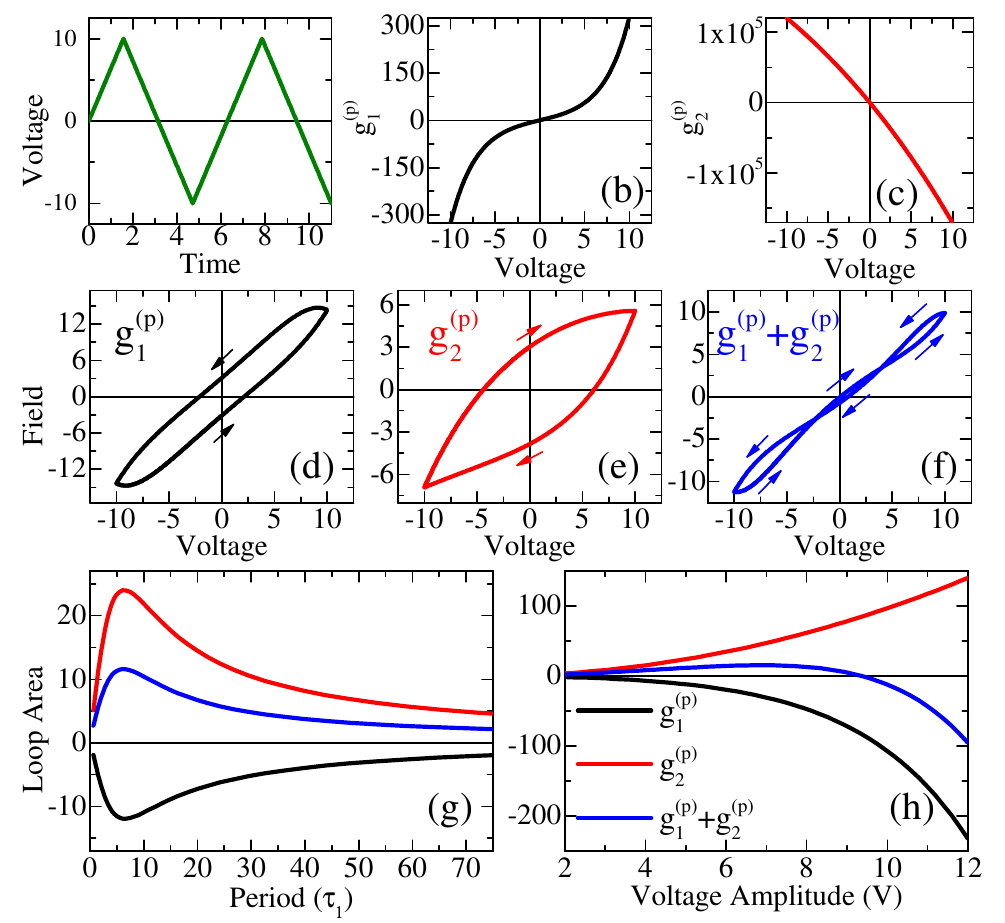}
    \caption{ (a) Periodic voltage input. Electric polarization transfer function for: (b) $g_1^{(p)}$, that increases nonequilibrium polarization with voltage and (c), the corresponding transfer function, $g_2^{(p)}$, for a leaking channel. Stable cycles of the electric field at the TMD layer in the presence of the process activated by: (d) $g_1^{(p)}$, (e) $g_2^{(p)}$, and (f) the combined effect of $g_1^{(p)}+g_2^{(p)}$ assuming equal relaxation times for both mechanisms and voltage sweeps with period $T= 2 \pi \tau_1$. The arrows in (d)-(f) indicate the direction of the $V$ sweep. (g) Voltage loop area as a function of the voltage period and (h) as a function of the voltage amplitude.}
    \label{fig2}
\end{figure}

Let us now discuss some particular cases of interest under periodic triangular voltage bias, $V$, with period T, as represented in Figure~\ref{fig2} (a). In this scenario, according to Eqs.~\ref{dp} and~\ref{dn1}, the system response will evolve towards a periodic behavior once any transient effects, which depend on the initial conditions, have faded. Note also that any dynamic process, $k$, for which $\tau_k << T$, will manifest itself as instantaneous and its contribution to the fluctuation of either polarization or charge will be reduced to $\delta P_k(V) = \tau_k g_k^{(p)}(V)$ or $\delta n_k(V) = \tau_k g_k^{(n)}(V)$, respectively. This reflects the nature of the transfer functions (or their superpositions) under very slow driving conditions. In turn, under adequate voltage drives, the stable cyclic response can reveal signatures of memory effects through hysteresis in the observable versus voltage characteristics. According to Eq.~\ref{ftmd3}, any hysteresis observed in the polarization versus $V$ will be reflected in any observable that depends on the local electric field at the TMD monolayer. For instance, reminiscent of ferroelectric apparent responses,~\cite{Scott2008} would manifest as counterclockwise loops. 

Two polarization transfer functions are illustrated in Figures~\ref{fig2}(b) and~\ref{fig2} (c). The first function results in an absolute increase in polarization for both positive and negative bias [Fig.~\ref{fig2}(b)], whereas the second function characterizes a potential dipole leakage with an opposite sign relative to the bias [Fig.~\ref{fig2}(c)]. Assuming the same relaxation times for each of these processes, $\tau_1=\tau_2$, and a voltage sweeping period of $T=2 \pi \tau_1$, the resulting local electric field hysteresis, according to Eqs.\ref{ftmd3} and \ref{dp}, is shown in Figures~\ref{fig2}(d) and~\ref{fig2} (e), respectively. Note that the nature of generation or leakage determines the direction of the hysteresis loop: counterclockwise loops for delayed increased polarization and clockwise loops if prevailing leaking polarization mechanisms~\cite{SilvaIgor2024}. In cases where these processes interfere, the loop's shape transforms into a multicrossing pattern, as depicted in Figure~\ref{fig2}(f). Due to the reactive nature of the delayed polarization response, the field hysteresis will consistently appear as open loops around $V=0$.
These loops collapse during very slow or very fast voltage sweeps using $\tau_1$ as reference as depicted in Figure~\ref{fig2}(g) by the calculated loop area. Discussions on the conditions for maximal memory response can be found in Refs.~\citenum{Paiva2022,LopezRichard2022}. The voltage amplitude also influences the shape of the loops, as shown in Figure~\ref{fig2}(g), where a change in the sign of the loop area indicates an inversion of the loop direction.

\begin{figure}[!htb]
    \centering
   \includegraphics[width=0.7\textwidth]{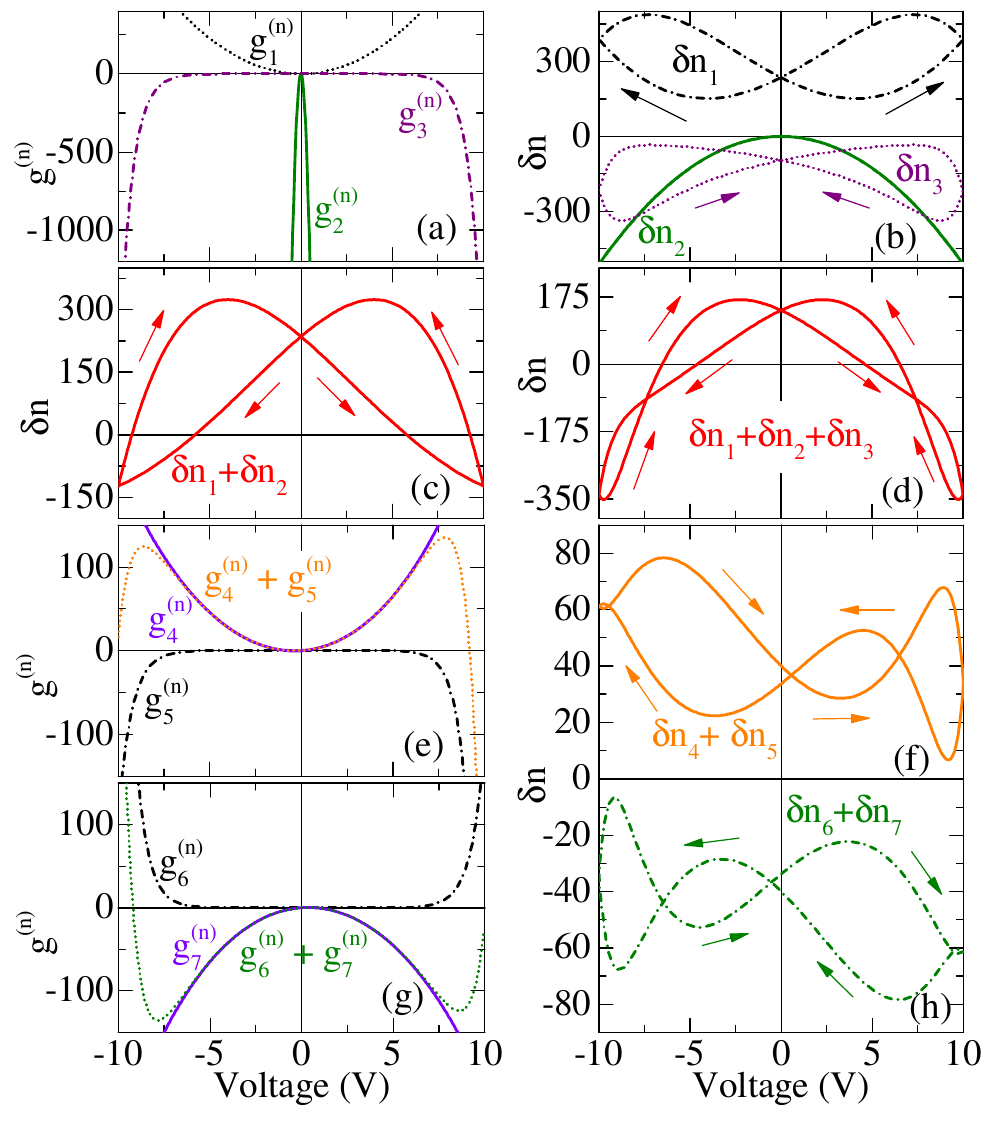}
    \caption{(a) Symmetric transffer functions for nonequilibrium charges of generation (positive) and trapping character (negative). (b) The corresponding loops for each of these three processes with $\tau_1=\tau_3>>\tau_2$ for voltage sweeps with period $T= 2 \pi \tau_1$. (c) Combined effect of the $\delta n_1+\delta n_2$ dynamics. (d) Combining $\delta n_1+\delta n_2+\delta n_3$. (e) Two asymmetric transfer functions with contrasting characters and efficiencies and the combineed effect of them displayed in panel (f). Panels (g) and (h) unveil the combination of opposite characters analogous to (e) and (f), respectively. The arrows indicate the orientation of the $V$ sweep.}
    \label{fig3}
\end{figure}

The charge fluctuations can exhibit a wide variety of responses, as illustrated in Fig.~\ref{fig3}. Panel~\ref{fig3} (a) presents three symmetric yet contrasting transfer functions that drive the charge fluctuations shown in panel~\ref{fig3} (b) under voltage cycles with $T= 2 \pi \tau_1$, with $\tau_3=\tau_1>> \tau_2$.
In these conditions, $\delta n_1$ and $\delta n_3$ exhibit pinched hysteresis at $V=0$ with opposite directions, while $\delta n_2\simeq \tau_2 g^{(n)}_2$, following a quasi-stationary (instantaneous) response, with no open hysteresis. As stated previously, it is possible to encounter concomitant processes with contrasting transfer functions and relaxation times. To illustrate this, we show the combined effects of adding $\delta n_1 +\delta n_2$, in panel~\ref{fig3} (c) and $\delta n_1 +\delta n_2+\delta n_3$ in panel~\ref{fig3} (d).

The histeresis crossing at $V=0$ is disrupted when inversion symmetry is broken. In Figure~\ref{fig3}, we illustrate the combined effect of two transfer functions with contrasting characteristics. In this scenario, the asymmetry is controlled by a single parameter in the transfer function, as detailed in Ref.\citenum{LopezRichard2022}. The symmetry break alters the loop's shape, as shown in Figure~\ref{fig3}(f), with the emergence of multiple loop intersections. These crossings can be attributed to the multiple sign changes of the function $g^{(n)}_4+g^{(n)}_5$ in panel~\ref{fig3} (e), although the positions of the intersections and the character change of the transfer functions (sign inversion with applied bias) are not correlated. 

By modifying the character and symmetry constraints of the transfer function, as displayed in panel~\ref{fig3} (g), the topology of the charge fluctuation changes, as represented in panel~\ref{fig3} (h), resulting in a complete inversion of the loop direction. Thus, the hysteresis shape and direction encapsulate crucial information about the nature of the internal microscopic mechanisms that lead to charge or polarization fluctuation.

Note that in our proposed Mem-emitter architecture, all internal variables are contained within the substrate material, while all observables emerge from the TMD monolayer. We must now correlate these internal variables with the memory traces of the observables of our Mem-emitter: the emission intensity Q and the position of the emission line at $\hbar \omega=\Delta E_{f,i}$ according to the approximations introduced in Eqs.~\ref{q} and~\ref{tau}.

In devices with inversion symetry break such as the one proposed in Figure~\ref{fig1} (e), the modulation of the optical transition rate in the 2D TMD monolayer can be proportional to the local electric field~\cite{Feng2019,Liu2022}
\begin{equation}
   \delta K_{f,i} \propto F_{TMD} (V).
    \label{dk}
\end{equation}
This is also the case for the renormalization of the effective energy gap in excitonic complexes under inversion asymmetry conditions~\cite{Shi2019,Feng2019}
\begin{equation}
   \delta \Delta E_{f,i} \propto F_{TMD} (V).
    \label{de}
\end{equation}
We should also note that a perpendicular electric field can result in different slopes for the renormalizations of the exciton and trion energies, as outlined in Ref.~\citenum{Borghardt2020}. 

Considering the optical transition rate (Eq.~\ref{tau})  $1/\tau=K_{f,i}(F_{TMD})\cdot D\left[ \Delta E_{f,i}(F_{TMD}) - \hbar \omega\right]$ , we expect the electric field memory modulation with the internal variables to be translated into a transition-rate-driven Mem-emitter, following the trends reported in Figure~\ref{fig2}, which could tune any excitonic complex emission. However, in the presence of electron fluctuation, $\delta n$, the trion emission intensity, $Q_{trion}$, has been proven to decouple from the exciton emission, $Q_{exc}$, given the expected ratio~\cite{Astakhov2002,Lundt2018}
\begin{equation}
   \frac{Q_{trion}}{Q_{exc}} \propto \left[ n_0+ \delta n\left(V \right) \right].
    \label{de}
\end{equation}
If the effect of electron fluctuation prevails over electric field modulation, we can anticipate excitons to be prone to exhibit transition rate-driven character in response to the local field, $F_{TMD}(V)$ as shown in Figure~\ref{fig2}. Conversely, trions may demonstrate a mixed response, acting as population-driven Mem-emitters, in line with the behavior of $\delta n(V)$ depicted in Figure~\ref{fig3}, while also showing transition rate-driven characteristics, similar to excitons. 
We believe that this control of stable light intensity and photon energy patterns can encode diverse information beyond the internal state characterization, with potential impact on communication and information processing.

In summary, the concept of Mem-emitters introduces an approach to enhancing and diversifying optoelectronic device functionalities.
In this perspective, understanding the differences between population-driven and transition-rate-driven Mem-emitters opens up a wide array of applications. Population-driven Mem-emitters offer the advantage of straightforward modulation of emitter density, making them suitable for optoelectronic switches, variable light sources, and applications where the control of active emitters is essential. In contrast, transition-rate-driven Mem-emitters provide refined control over emission properties, making them ideal for high-precision and advanced optoelectronic applications, such as communication systems and sensors requiring precise emission control. By exploring these contrasting mechanisms, and the information encoded in the light emission patterns, it is possible to develop more versatile and efficient light-emitting devices tailored to specific needs in various technological fields. Based on this analysis, two conditions are sufficient for Mem-emitter abilities to emerge (or be resolved): (i) the proper tuning of external drives to match the internal time scale appropriate for a given internal state variable [see Fig.~\ref{fig2} (g)] and (ii) the comparable strength of fluctuations of these variables with respect to their equilibrium or instantaneous reference values, according to Eqs.~\ref{p} and~\ref{de}. In particular, we anticipate that these unique properties could pave the way for the development of innovative technologies for memory and computation. TMDs and other 2D semiconductor monolayers offer a versatile and powerful platform for developing Mem-emitter functionalities when coupled to adequate substrates due to their unique properties, including atomic-scale thickness, tunable electronic properties, and strong light-matter interaction. Their ability to be modulated by external fields and integrated with other materials enhances their potential for advanced optoelectronic applications and in particular those related to memory effects. This underscores the critical importance of selecting appropriate substrates that would provide the ingredients for memory-related dynamics. Given the ubiquitous nature of the memory effects described, we can foresee that mem-emitting properties can also serve as a versatile tool for characterizing the intrinsic dynamics of complex systems beyond the functionalities of the Mem-emitter itself.
\begin{acknowledgement}
This study was financed in part by the Coordenação de Aperfeiçoamento de Pessoal de Nível Superior - Brazil (CAPES) and the Conselho Nacional de Desenvolvimento Científico e Tecnológico - Brazil (CNPq) Projs. 311536/2022-0, 309920/2021-3, and 306170/2023-0.
\end{acknowledgement}

%\bibliography{References}
\bibliography{Mememitters.bbl}
\end{document}